\documentclass[aps,amsmath,amssymb,prx,reprint,groupedaddress]{revtex4-2}

\usepackage{graphicx}
\usepackage{dcolumn}
\usepackage{bm}
\usepackage{hyperref}
\hypersetup{colorlinks=true,citecolor=blue,urlcolor=blue,linkcolor=blue}

\newcommand{\suppref}{SM \footnotemark[1]}

\begin{document}


\title{Quasinormal mode theory for nanoscale electromagnetism with quantum surface responses}

\author{Qiang Zhou}
\affiliation{School of physics and Wuhan National Laboratory for Optoelectronics, Huazhong Univeristy of Science and Technology, Luoyu Road 1037, Wuhan, 430074, People's Republic of China}

\author{Pu Zhang}
\email[Corresponding auther: ]{puzhang0702@hust.edu.cn}
\affiliation{School of physics and Wuhan National Laboratory for Optoelectronics, Huazhong Univeristy of Science and Technology, Luoyu Road 1037, Wuhan, 430074, People's Republic of China}

\author{Xue-Wen Chen}
\email[Corresponding auther: ]{xuewen\_chen@hust.edu.cn}
\affiliation{School of physics and Wuhan National Laboratory for Optoelectronics, Huazhong Univeristy of Science and Technology, Luoyu Road 1037, Wuhan, 430074, People's Republic of China}

\begin{abstract}
We report a self-consistent quasinormal mode theory for nanometer scale electromagnetism where the possible nonlocal and quantum effects are treated through quantum surface responses. With Feibelman’s frequency-dependent \textit{d} parameters to describe the quantum surface responses, we formulate the source-free Maxwell’s equations into a generalized linear eigenvalue problem to define the quasinormal modes. We then construct an orthonormal relation for the modes and consequently unlock the powerful toolbox of modal analysis. The orthonormal relation is validated by the reconstruction of the full numerical results through modal contributions. Significant changes in the landscape of the modes are observed due to the incorporation of the quantum surface responses for a number of nanostructures. Our semi-analytical modal analysis enables transparent physical interpretation of the spontaneous emission enhancement of a dipolar emitter as well as the near-field and far-field responses of planewave excitations in the nanostructures.
\\
\end{abstract}

\maketitle

\section{Introduction}
Nano-optics has flourished with the developments of new concepts in optical physics and emergence of light-enabled nanotechnologies. Light confinement, as a key enabler, has well surpassed the diffraction limit and entered nanometer length scale \cite{Ebbesen2003SurfacePlasmons,Brongersma2010Plasmonics,Novotny2012NanoOptics,Polman2015Nanophotonics,Sandoghdar2020}.
Surface plasmons supported by metal nanostructures are the workhorse for light confinement \cite{Ebbesen2003SurfacePlasmons,Brongersma2010Plasmonics}.
Multiscale plasmonic systems with nanoscopic features (\textit{e.g.}, nanogaps) in a mesoscopic host metal structure \cite{Baumberg2019ExtremeNanophotonics}, have attracted considerable interest, as they can reveal the nanoscopic details through far-field spectroscopic measurements.
Recent advances in nanotechnologies have demonstrated the fabrication of ultra-fine nanostructures and meticulous nano-manipulation of the minuscule parts, such that making features with sizes down to 1 nm and even smaller becomes feasible  \cite{Norris2009,Cirac2012Probing,Dionne2012QuantumPlasmon,Hecht2012AtomicScale,Baumberg2015}.
In nanometer length scale, non-classical effects including electron nonlocality \cite{Cirac2012Probing,Stenger2015}, spill-over \cite{Hongxing2015ResonanceShifts,Ding2017Plasmonic} and Landau damping \cite{cirac2017CurrentDependent,Khurgin2017LandauDamping} start to play significant roles. Calculations on field enhancement \cite{Khurgin2017LandauDamping}, spontaneous emission rate \cite{Rockstuhl2014}, hot electron generation \cite{Govorov2017}, plasmon resonances \cite{Dionne2012QuantumPlasmon,Cirac2012Probing} and spasing \cite{Gamacharige2019}, to name a few, all should include the possible non-classical effects. First-principles calculations such as time-dependent density functional theory are in principle exact, and has been used in a number of proof-of-principle studies \cite{Borisov2012,Townsend2011,Zuloaga2009QuantumDescription,Rossi2015,Morton2011}. However these studies are restricted to metal clusters or systems of at most few-nanometer size, because the computational cost quickly grows intractable for multiscale systems. A quantum surface response (QSR) description based on Feibelman's $d$-parameters stands out as an efficient approach to accounting for major non-classical effects \cite{Feibelman1982}. In recent years, the $d$-parameter description has rapidly developed and become increasingly sophisticated \cite{Yan2015,Shvonski2017,Christensen2017,Yang2018,Kong2018,Yang2019NatureFramework,Echarri2020,PAD2020,PAD2020b}. In particular, it culminated in a general theoretical framework for nano-optics \cite{Yang2019NatureFramework}, which proves effective for covering surface bound non-classical effects, \textit{e.g.} nonlocality and Landau damping \cite{PAD2020}.

While QSRs with the $d$-parameter description can be incorporated into calculations, transparent interpretation of the involved optical processes is difficult with the brute force numerics. A majority of optical processes in nano-optics often can be conceptualized with a few resonant modes. Therefore a mode theory compatible with the $d$-parameter description is of key importance and would facilitate the establishment of a general theoretical framework for nanometer scale electromagnetism. In view of the openness and dissipative nature of plasmonic systems, here a quasinormal mode (QNM) theory is necessitated. The classical QNM theory in fact has been available and well received \cite{Lalanne2013PRLTheory,Lalanne2018LigntInteraction,YanWei2018PRB,Binkowski2018RieszProjection,Muljarov2019ResonantExpansion,Philip2020biorthogonalQNM}.
Various aspects of nano-optical theory including scattering \cite{Kuipers2017}, harmonic generation \cite{Gigli2020}, coupled mode theory \cite{Miller2020,Tali2020}, perturbation theory \cite{Qiu2020,Alegre2020}, photon-emitter interaction \cite{Kewes2018StrongCoupling,Gurlek2018} and field quantization \cite{Young1998,Richter2019} were revisited in the context of QNM theory. QNM theory has also been generalized to incorporate non-classical effects based on volume-wise material description \cite{Stephen2017NonlocalQNM,Binkowski2019ModalAnalysis,Zhou2021}.
While a perturbative $d$-parameter correction to the QNM eigenfrequency was discussed in Refs.\,\cite{Christensen2017,Yang2019NatureFramework}, a non-perturbative QSR informed QNM theory remains elusive. More importantly, the orthonormalization of the QNMs compatible with $d$-parameter description is an open question. QNM orthonormalization is inherently challenging \cite{Kristensen2015,Langbein2017} and at the same time a crucial ingredient of a QNM theory, since only with the orthonormal relation one can quantitatively and analytically describe the optical responses with the modal contributions.

Here we develop a non-perturbative QNM theory for nanoscale electromagnetism with the incorporation of QSRs. The source-free Maxwell's equations incorporated with the QSRs based on $d$-parameter description are formulated into a volume-surface composite linear eigenvalue problem (LEVP). Moreover, we show a route to obtain an orthonormal relation for the QNMs which directly leads to the proper QNM normalization. The orthonormal relation, together with suitably designated formal source terms, further empowers analytical expansion of system responses in terms of QNMs. Lastly we demonstrate that the full-fledged mode theory provides physically transparent interpretations of non-classical optical responses to dipole emission and to the far-field excitation.

\begin{figure}[tb!]
\centering
\includegraphics[width=0.45\textwidth]{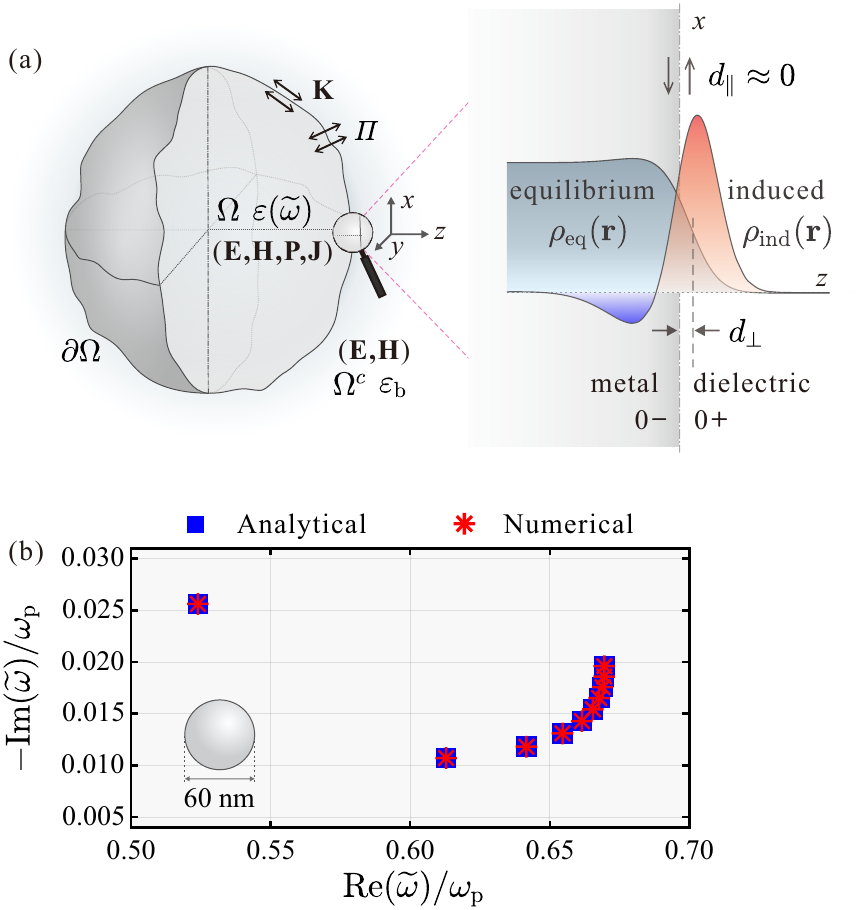}
\caption{(a) The quantum surface responses (QSRs) of a metallic nanostructure in terms of effective surface current $\mathbf{K}$ and polarization $\varPi$. The zoomed-in view illustrates the QSRs characterized by Feibelman's $d$ parameters at the metal-dielectric surface. (b) Principal QNM eigenfrequencies of a sodium ($r_s=4$, $\hbar\gamma=0.1$ eV) nanosphere in vacuum obtained numerically by solving the generalized linear eigenvalue problem Eq.\,(\ref{Eq:LEVP1},\,\ref{Eq:LEVP2}) and analytically by locating the poles of the generalized Mie coefficients \cite{PAD2020}. $d_\perp$ is taken from Fig.\,\ref{Fig:2}\,(a).}
\label{Fig:1}
\end{figure}

\section{QSR informed QNM theory}
Quantum surface responses originate from microscopic eletron dynamics bounded at metal surfaces.
Assuming metal surfaces are microscopically flat, we look locally at the structure surface \textit{e.g.} $z=0$ as illustrated in Fig.\,\ref{Fig:1}\,(a).
A charge distribution $\rho_\mathrm{ind}(\mathbf{r})$ is induced around the surface by an optical excitation.
The QSRs on the nanoparticle then manifest macroscopically as an effective surface current $\mathbf{K}$ and surface polarization $\varPi$ resulting from $\rho_\mathrm{ind}(\mathbf{r})$ and the corresponding current $\mathbf{J}_\mathrm{ind}(\mathbf{r})$.
Under $d$-parameter description, the finite extent of the light-induced current distribution $\mathbf{J}_\mathrm{ind}(\mathbf{r})$ is described on the leading order by \cite{Feibelman1982,Liebsch1997}
\begin{equation}
d_\perp = \frac{
\int_{-\infty}^{\infty}dz\,z\frac{d}{dz}J_{z,\mathrm{ind}}(z)}{
\int_{-\infty}^{\infty}dz\,\frac{d}{dz}J_{z,\mathrm{ind}}(z)},\quad
d_\parallel = \frac{
\int_{-\infty}^{\infty}dz\,z\frac{d}{dz}J_{\parallel,\mathrm{ind}}(z)}{
\int_{-\infty}^{\infty}dz\,\frac{d}{dz}J_{\parallel,\mathrm{ind}}(z)}.
\end{equation}
Then the effective surface current and polarization can be approximated as
\begin{equation}
\mathbf{K} = i\widetilde\omega d_\parallel[\![\mathbf{D}_\parallel]\!],\quad
\varPi = d_\perp\varepsilon_0[\![E_\perp]\!].
\end{equation}
In the above equations, the time convention $e^{-i\widetilde\omega t}$ is assumed. We have adopted $[\![f]\!]\equiv f(0+)-f(0-)$ and the notation $\perp$/$\parallel$ to denote the normal/parallel component of a vector.
$\mathbf{K}$ and $\varPi$ in turn couple back to Maxwell's equations as secondary sources and consistently modify the overall optical responses.
The main parameter $d_\perp$ can be rewritten as
$d_\perp =
\int_{-\infty}^{\infty}dz\,z\rho_{\mathrm{ind}}(z) /\allowbreak
\int_{-\infty}^{\infty}dz\,\rho_{\mathrm{ind}}(z)$
by using Gauss's law.
It thus appears as the centroid of $\rho_\mathrm{ind}(z)$, a characteristic length scale of the theory.
The other parameter $d_\parallel$ however turns out to be negligible in many situations \cite{Yang2019NatureFramework}.
Hence, we assume $d_\parallel=0$ in the following, and only the surface polarization remains
\begin{equation}
\varPi(\mathbf{r}_\parallel) 
= d_\perp[\![\varepsilon^{-1}]\!]\,D_\perp^{s}(\mathbf{r}_\parallel).
\end{equation} 
Here $\mathbf{r}_\parallel$ denotes the coordinates on the surface $\partial\Omega$, and $\varepsilon$ is relative permittivity.
$D_\perp^s$ corresponds to the electric displacement field $\mathbf{D}$ on $\partial\Omega$.
A brief overview of the $d$-parameter description is summarized in the Supplementary Material (SM) \footnotemark[1].
Noticeably $d$-parameter description has deeper implications that underlie its efficacy of modeling non-classical effects.
The fact that $d$-parameters are alternatively expressible with nonlocal permittivity \cite{Liebsch1997} rationalizes the ability to mimic nonlocal material.
Their frequency dispersion indicates nontrivial QSRs and the corresponding imaginary parts induce surface contribution to loss.

To set the stage for modal analysis under $d$-parameter description, having an LEVP formulation is conceptually important.
In this regard, the bulk material response inside the metal domain $\Omega$ is treated the same as the classical QNM theory \cite{YanWei2018PRB}.
Specifically we assume the metal is described by the Drude-Lorentz model $\varepsilon(\widetilde\omega)=\varepsilon_\infty-\omega_\mathrm{p}^2/(\widetilde\omega^2-\omega_0^2+i\gamma\widetilde\omega)$ with $\varepsilon_\infty$, $\omega_\text{p}$, $\omega_0$ and $\gamma$ being the background permittivity, plasma frequency, resonance frequency and damping rate, respectively.
The response is linearized regarding $\widetilde\omega$ by introducing the auxiliary polarization \textbf{P} and the corresponding current density \textbf{J}.
In general, the parameter $d_\perp$ is also dispersive regarding $\widetilde\omega$ as retrieved from \textit{e.g.}, TDDFT calculations and experiments.
We can similarly treat the QSRs characterized with $d_\perp$ by exploiting the pole structure of 
$f(\widetilde\omega) \equiv d_\perp(\widetilde\omega)[\![\varepsilon^{-1}(\widetilde\omega)]\!]$.
Considering that $f(\widetilde\omega)$ is bounded at $\widetilde\omega=0$ and $\infty$, we may expand it over the simple poles $\widetilde\omega_\beta$ as
\begin{equation}
f(\widetilde\omega)
= \sum_{\beta=0}^Nf_\beta\,\frac{\widetilde\omega}{\widetilde\omega-\widetilde\omega_\beta}.
\end{equation}
A pseudo-pole $\widetilde\omega_0=0$ is formally introduced for conciseness of expression.
Thereupon, we introduce for each pole an auxiliary surface polarization
$\varPi_{\beta}=\widetilde\omega D_\perp^s/(\widetilde\omega-\widetilde\omega_{\beta})$,
with the constraint 
$\sum_{\beta=0}^Nf_\beta\varPi_\beta=\varPi$ \footnotemark[1].

In terms of the auxiliary fields $\mathbf{P}$, $\mathbf{J}$ and $\varPi_\beta$, the source-free Maxwell's equations now become linearized.
The system can be formulated into a generalized LEVP comprising inter-coupled volume and surface components
\begin{gather}
\label{Eq:LEVP1}
\left[\begin{matrix}
0&\frac{i\nabla\times}{\varepsilon_0\varepsilon_\infty}&0&\frac{1}{i\varepsilon_0\varepsilon_\infty}\\
\frac{\nabla\times}{i\mu_0}&0&0&0\\
0&0&0&i\\
i\varepsilon_0\omega_\mathrm{p}^2&0&-i\omega_0^2&-i\gamma
\end{matrix} \right]
\widetilde\Phi_\mathrm{v}
= \widetilde{\omega}\,\widetilde\Phi_\mathrm{v}
+ \mathcal{V}_\mathrm{vs},
\\[5pt]
\label{Eq:LEVP2}
\mathrm{diag}\!\left\{\widetilde\omega_0,\widetilde\omega_1,\cdots,\widetilde\omega_N\right\}
\widetilde\Phi_\mathrm{s}
= \widetilde\omega\,\widetilde\Phi_\mathrm{s}
+ \mathcal{V}_\mathrm{sv},
\end{gather}
where $\varepsilon_0$, $\mu_0$ and $\widetilde\omega$ are vacuum permittivity, permeability and the complex eigenfrequency, respectively.
The composite eigenvector $\widetilde\Phi=\widetilde\Phi_\mathrm{v}\oplus\widetilde\Phi_\mathrm{s}$ consists of 
$\widetilde\Phi_\mathrm{v} = [\mathbf{\widetilde E},\mathbf{\widetilde H},\mathbf{\widetilde P},\mathbf{\widetilde J}]^\mathrm{T}$ in bulk domains $\Omega+\Omega^c$ 
and 
$\widetilde\Phi_\mathrm{s} = [\widetilde\varPi_0,\widetilde\varPi_1,\cdots,\widetilde\varPi_N]^\mathrm{T}$ on $\partial\Omega$.
Here $\Omega^c$ is the complement of $\Omega$.
The supporting domains of $(\mathbf{\widetilde E},\mathbf{\widetilde H})$ and $(\mathbf{\widetilde P},\mathbf{\widetilde J})$ in $\widetilde\Phi_\mathrm{v}$ shall be understood separately.
The inter-coupling sources that couple $\widetilde\Phi_\mathrm{v}$ and $\widetilde\Phi_\mathrm{s}$ are
$\mathcal{V}_\mathrm{vs} = [\frac{\widetilde\omega\widetilde\varPi}{\varepsilon_0\varepsilon_\infty},0,0,0]^\mathrm{T}$ and 
$\mathcal{V}_\mathrm{sv} = -\widetilde\omega\widetilde{D}_\perp^s[1,1,\cdots,1]^\mathrm{T}$.
Note that $\widetilde D_\perp^s$ serves as the field mediating the two components.
In addition, the radiation boundary condition should be respected in the far field.

\begin{figure*}[htb!]
\centering
\includegraphics[width=0.9\textwidth]{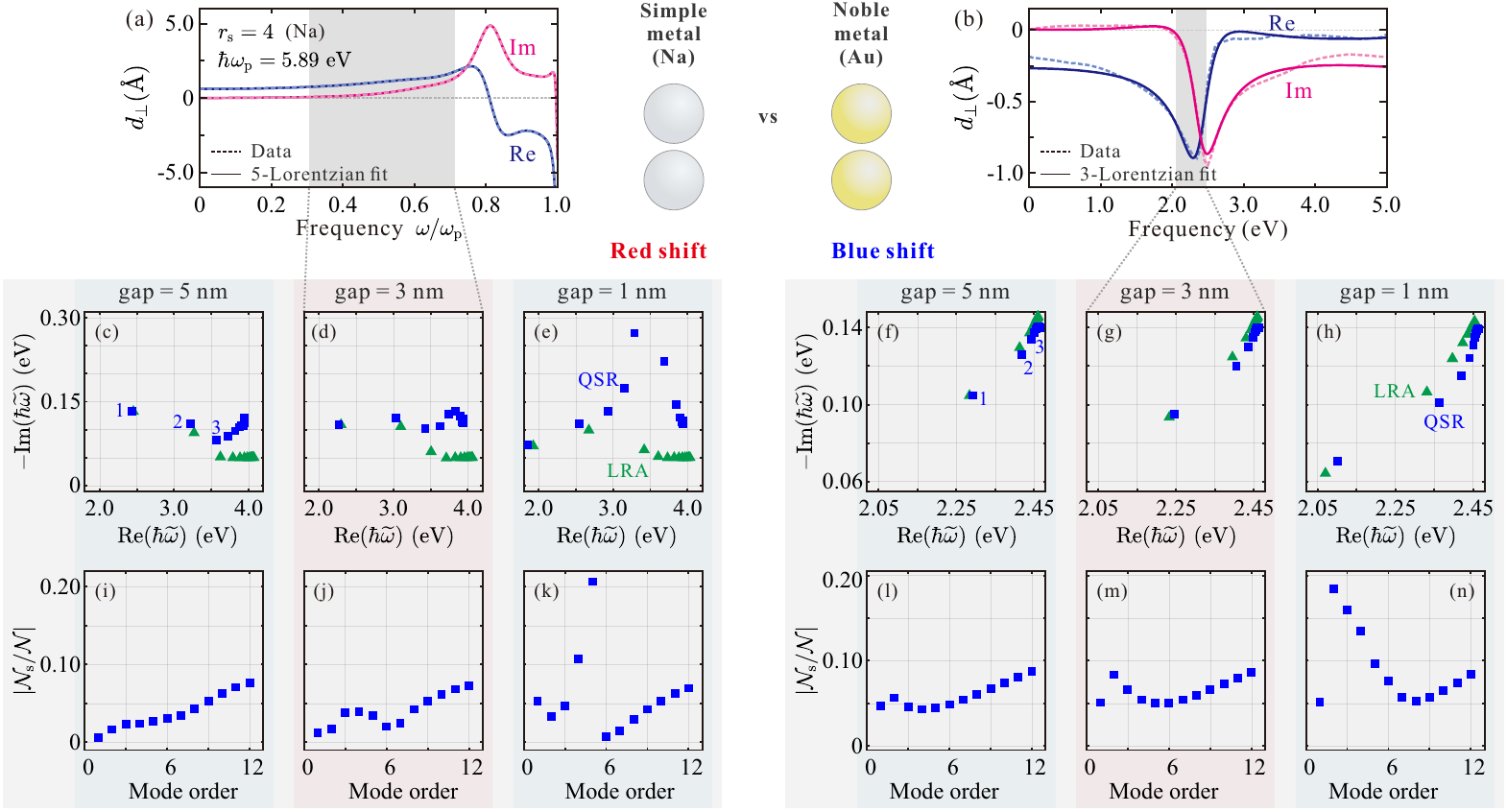}
\caption{The QSR informed QNMs of sodium (left panel) and gold (right pannel) nanosphere dimers.
The sphere diameter is 60 nm.
(a, b) The Feibelman's $d_\perp$ parameters for the metal-air interfaces calculated with high-level theories (TDDFT for sodium \cite{Christensen2017} and specular reflection model for gold \cite{Echarri2020}) and the multiple Lorentzian fittings.
The fitting for sodium is according to Ref.\,\cite{PAD2020}.
(c-h) The first 12 QNM eigenfrequencies for various gap sizes calculated with the QSRs (blue squares) and under LRA (green triangles).
(i-m) The ratios between the surface ($\mathcal{N}_\mathrm{s}$) and the complete ($\mathcal{N}$) QNM normalization factors of the first 12 QNMs for various gap sizes.}
\label{Fig:2}
\end{figure*}

The eigenmodes or QNMs defined above can be numerically solved based on a quadratic reformulation of the LEVP and a boundary PDE implementation for $d_\perp(\widetilde\omega)$, which is fitted with multiple Lorentzians \footnotemark[1].
As a benchmark, the first $10$ QNM eigenfrequencies of a sodium nanosphere is solved and compared with analytical results.
Sodium is modeled as an electron gas with a Wigner-Seitz radius of $r_\mathrm{s}=4$.
The decay rate is $\hbar\gamma=0.1$ eV and $d_\perp(\widetilde\omega)$ adopts the fitting presented in Fig.\,\ref{Fig:2}(a).
$\widetilde\omega_m$ are obtained analytically by locating the poles of the generalized Mie coefficients \cite{PAD2020}.
As shown in Fig.\,\ref{Fig:1}\,(b), our numerical implementation gives accurate results.

\section{Orthonormalization of QSR informed QNM}

The normalization of the resulting QNMs of the LEVP Eqs.\,(\ref{Eq:LEVP1},\,\ref{Eq:LEVP2}) is an indispensable part of a complete QNM theory.
However, the QNM profiles have the issue of exponential divergence in the far field as their classical counterparts owing to the non-Hermitian nature of the LEVP \cite{Lalanne2013PRLTheory,Lalanne2013OE_IntuitiveMethod}.
The issue makes the normalization a difficult problem to tackle \cite{Lalanne2013PRLTheory}.
The monotonic divergence appearing in the sesquilinear form commonly used for the normalization of normal modes can be converted by using a bilinear form into an oscillatory diverging quantity \cite{Leung1994,Lalanne2013PRLTheory,Kristensen2015,Langbein2016}.
Its integral then becomes managable with the technique of complex coordinate transformation and produces physically meaningful value \cite{Lalanne2013PRLTheory}.
Normalization based on bilinear form was first applied to classical local media and has recently been generalized by the authors to general nonlocal media \cite{Zhou2021}.
Nevertheless, when non-classical effects exclusively occur at surfaces, $d$-parameter description is more favorable.
For QSR informed QNMs the bilinear forms accounting for only material response in bulk domains are obviously inapplicable.

Here we recognize the contribution of the QSRs and provide a procedure to properly orthonormalize the QNMs based on generalized unconjugated Lorentz reciprocity \footnotemark[1].
In particular, an extra surface integral of $\widetilde\varPi_\beta$ over $\partial\Omega$ appears in the bilinear form as the QSR contribution.
Consequently we arrive at a volume-surface composite bilinear form $(\!(\widetilde{\Phi}_m,\widetilde{\Phi}_n)\!)_\mathrm{c}\equiv (\!(\widetilde{\Phi}_{m,\mathrm{v}},\widetilde{\Phi}_{n,\mathrm{v}})\!)_\mathrm{v}+(\!(\widetilde{\Phi}_{m,\mathrm{s}},\widetilde{\Phi}_{n,\mathrm{s}})\!)_\mathrm{s}$ (see \suppref{} for the derivation),
\begin{align}
(\!(\widetilde\Phi_{m,\mathrm{v}},\widetilde\Phi_{n,,\mathrm{v}})\!)_\mathrm{v}
&= \int{d}\mathrm{r}\,\Big\{
\varepsilon_0\varepsilon_\infty\mathbf{\widetilde E}_m\cdot\mathbf{\widetilde E}_n
- \mu_0\mathbf{\widetilde H}_m\cdot\mathbf{\widetilde H}_n
\notag\\
&\
+ \frac{\omega_0^2}{\varepsilon_0\omega_\mathrm{p}^2}\mathbf{\widetilde P}_m\cdot\mathbf{\widetilde P}_n
- \frac{1}{\varepsilon_0\omega_\mathrm{p}^2}\mathbf{\widetilde J}_m\cdot\mathbf{\widetilde J}_n
\Big\},
\\
\label{Eq:compBF}
(\!(\widetilde{\Phi}_{m,\mathrm{s}},\widetilde{\Phi}_{n,\mathrm{s}})\!)_\mathrm{s}
&=
-\int_{\partial\Omega}d\mathbf{r}_\parallel 
\sum^N_{\beta=0} \frac{f_\beta}{\varepsilon_0}\,
\widetilde{\varPi}_{m,\beta}\widetilde{\varPi}_{n,\beta}.
\end{align}
The volume part remains unchanged and identical to the bilinear form in classical QNM theory \cite{YanWei2018PRB}.
The composite bilinear form immediately leads to the orthogonal relation 
$(\!(\widetilde{\Phi}_m,\widetilde{\Phi}_n)\!)_\mathrm{c}=0$ 
when $\widetilde\omega_m\neq \widetilde\omega_n$.
For a nondegenerate QNM $\widetilde\Phi'_m$, it can be normlized as 
$\widetilde\Phi_m=\widetilde\Phi'_m/\sqrt{\mathcal{N}_m}$.
Naturally the normalization factor
$\mathcal{N}_m=(\!(\widetilde{\Phi}'_m,\widetilde{\Phi}'_m)\!)_\mathrm{c}
\equiv \mathcal{N}_{m,\mathrm{v}}+\mathcal{N}_{m,s}$
as well consists of the unchanged classical one \cite{Lalanne2018LigntInteraction} and a surface contribution,
\begin{align}\label{Eq:N_ms}
\mathcal{N}_{m,\mathrm{v}}
&= \int{d}\mathbf{r}
\left(2+\widetilde\omega_m\frac{\partial}{\partial\widetilde\omega_m}\right)
\varepsilon(\widetilde\omega_m)\,
\varepsilon_0\mathbf{\widetilde E}_m\cdot\mathbf{\widetilde E}_m,
\\
\mathcal{N}_{m,\mathrm{s}}
&= -\int_{\partial\Omega}d\mathbf{r}_\parallel
\left(2-\widetilde\omega_m\frac{\partial}{\partial\widetilde\omega_m}\right)
d_\perp(\widetilde\omega_m)[\![\varepsilon^{-1}(\widetilde\omega_m)]\!]
\notag\\
&\qquad\qquad
\varepsilon_0^{-1}\,\widetilde D_{m,\perp}\widetilde D_{m,\perp}.
\end{align}
The QNMs are assumed normalized hereafter.
For the normalized QNMs the electric complex mode volume is defined as
\begin{equation}
\widetilde V_m(\mathbf{r})
= \frac{1}{2\varepsilon_0n_d^2(\mathbf{r})\widetilde{\mathbf{E}}_m^2(\mathbf{r})},
\end{equation}
where $n_d(\mathbf{r})$ is refractive index.
The real-valued effective mode volume is $V_{\mathrm{eff}}\equiv1/\mathrm{Re}[\widetilde{V}_m^{-1}(\mathbf{r})]$.

The normalized QNMs and their orthogonal relation comprise the cornerstone of the QSR informed QNM theory.
The orthogonal relation specially empowers the semi-analytical expansion of the optical responses of a nanosystem under an arbitrary excitation.
The linear response to a weak excitation at real frequency $\omega$ can be expanded as 
$\Psi(\omega)=\sum_m\alpha_m(\omega)\,\widetilde\Phi_m$.
Applying the orthogonal relation, the expansion coefficients are derived analytically as \footnotemark[1]
\begin{equation}
\label{QNMexpansion}
\alpha_m(\omega)=\frac{1}{\widetilde\omega_m-\omega}(\!(\widetilde\Phi_m,\mathcal{S})\!)_\mathrm{c}.
\end{equation}
The composite formal source $\mathcal{S}=\mathcal{S}_\mathrm{v}\oplus\mathcal{S}_\mathrm{s}$ contains a volume component $\mathcal{S}_\mathrm{v}$ and a surface component $\mathcal{S}_\mathrm{s}$, and assumes an excitation dependent form.
For a current source $\mathbf{J_s}$, $\mathcal{S}$ only has a nonzero volume component
\begin{equation}
\label{DipoleSource}
\mathcal{S}_\mathrm{v} = [i\mathbf{J_s}/(\varepsilon_0\varepsilon_\infty),\allowbreak 0,0,0]^\mathrm{T}.
\end{equation}
As a special case, a dipolar source of dipole moment \textbf{d} at $\mathbf{r}_\mathrm{d}$ has $\mathbf{J_s}=-i\omega\mathbf{d}\delta(\mathbf{r}-\mathbf{r}_\mathrm{d})$.
The expansion coefficient correspondingly simplifies to
$\alpha_m=-\omega\widetilde{\mathbf{E}}_m(\mathbf{r}_\mathrm{d})\cdot\mathbf{d}/(\omega-\widetilde\omega_m)$.
A QNM's contribution to Purcell factor can be defined as
$f_{\mathrm{P},m}\equiv P_m/P_0=-\int d\mathbf{r}\mathrm{Re}\{\mathbf{J}^*_\mathrm{s}\cdot\mathbf{\widetilde{E}}_m\}/(2P_0)$
with $P_0$ being the power radiated by the dipole in vacuum. $P_m$ and $\mathbf{\widetilde{E}}_m$ are the modal contributions to the radiated power and electric field, respectively.
Then the modal Purcell factor turns out to be
\begin{equation}
\label{Eq:fPm}
f_{\mathrm{P},m}=\frac{3\pi c^3}{n^2_\mathrm{d}\omega^2}\mathrm{Im}\Big\{\frac{1/V_m}{\widetilde{\omega}_m-\omega}\Big\}.
\end{equation}
For an incident field $\mathbf{E}_\mathrm{inc}$, the formal source is expressed with
\begin{align}
\label{PWSource}
\mathcal{S}_\mathrm{v}&= [\,(\Delta\varepsilon_\infty/\varepsilon_\infty)\,\omega\mathbf{E}_\mathrm{inc},\allowbreak 0,0,-i\varepsilon_0\omega_\mathrm{p}^2\mathbf{E}_\mathrm{inc}]^\mathrm{T},\\
\mathcal{S}_\mathrm{s}&= -\omega\varepsilon_0\varepsilon_\mathrm{b}\mathbf{E}_\mathrm{inc,\perp}[1,1,\cdots,1]^\mathrm{T}.
\end{align}
Here $\Delta\varepsilon_\infty=\varepsilon_\infty-\varepsilon_\mathrm{b}$
with $\varepsilon_\mathrm{b}$ being the relative permittivity of the scattering background.
Using the response expansion in \eqref{QNMexpansion} extinction cross section could as well be decomposed into QNM contributions.
Noteworthily, the evaluation and expansion of the Purcell and extinction spectra make use of the modified Poynting theorem under $d$-parameter description as detailed in \suppref{}.


\section{QSR influence on optical eigenmodes}
The eigenmodes of a nanoplasmonic system often play a central role in the optical responses.
Here we first scrutinize how the QSRs influence the eigenmodes.
As a prototypical motif in nanoplasmonics, we consider metal nanosphere dimers as examples.
The conduction electrons in noble metals are known to behave distinctly from those in simple metals due to bound electron screening.
Dimers made of sodium and gold are thus examined in parallel as shown in the left and right panels of Fig.\,\ref{Fig:2}.
The complex $d_\perp$ parameters of Na and Au are plotted in Fig.\,\ref{Fig:2}(a) and \ref{Fig:2}(b), respectively.
The dashed lines in Fig.\,\ref{Fig:2}(a) for Na are obtained from TDDFT calculations \cite{Christensen2017} while the dashed traces for Au in Fig.\,\ref{Fig:2}(b) are retrieved from the specular reflection model \cite{Echarri2020}.
The solid traces in Fig.\,\ref{Fig:2}(a) and \ref{Fig:2}(b) are the multiple Lorentzian fittings of the corresponding dashed lines (see \suppref{} for the material parameters).
The dimers' QNMs under study lie in the frequency ranges shaded in gray, where the $d_\perp$ parameters of the two metals show striking differences in several aspects.
While $d_\perp$ is weakly dependent on frequency for Na, strongly dispersive $d_\perp$ is observed for Au.
Moreover Na has larger positive $\mathrm{Re}\{d_\perp\}$, characteristic of considerable electron spill out of the metal boundary.
For Au, $\mathrm{Re}\{d_\perp\}$ takes negative values and the magnitude is smaller, which is consistent with typical slight electron spill-in for noble metals.

To examine the influence of $d_\perp$ parameters on the eigenmodes, we perform the QNM analysis for the Na and Au dimers with 5 nm, 3 nm and 1 nm gaps.
The resulting QNM spectra for all the cases are summarized in Fig.\,\ref{Fig:2}(c)-\ref{Fig:2}(h). Since the dimer structure and dipole exictation concerned in later discussion both have cylindrical symmetry, only the QNMs of azimuthal order $m=0$ are presented. The eigenfrequencies of the QNMs calculated based on classical LRA are displayed with green triangle markers while the results of the QSR informed QNM calculations are depicted with blue squares.
The immediate observation is that the QSR informed QNMs are redshifted with respect to the classical QNMs for Na dimers while they get blueshifted for Au dimers. It in fact reflects the opposite signs of $\mathrm{Re}\{d_\perp\}$ for Na and Au. Positive/negative $\mathrm{Re}\{d_\perp\}$ leads to redshift/blueshift of the eigenmodes. Besides the real parts, the imaginary parts of the eigenfrequencies also display interesting behaviors.
The positive $\mathrm{Im}\{d_\perp\}$ of Na causes more damping in the QNMs, yet the negative $\mathrm{Im}\{d_\perp\}$ of Au seems to mitigate some dissipation.

When inspecting the variations of the modal properties against gap size and mode order, Na and Au dimers instead show some common features.
Most of the time the corrections to the LRA predictions, no matter in the real or imaginary parts of the eigenfrequencies, are greater for smaller gaps and higher-order modes.
For both cases, the QSRs gain more weight as the QNMs have more confined field near metal surface and experience stronger QSRs. The significance of the QSRs in a QNM could be partially quantified with an indicator $|\mathcal{N}_{\mathrm{s}}/\mathcal{N}|$.
The ratio is plotted in Fig.\,\ref{Fig:2}(i)-\ref{Fig:2}(n) for all the dimers as functions of the mode order.
Although the ratios show non-monotonic dependence on mode order, the up to 20\% values undoubtedly unveil the generally significant effects of the QSRs on the eigenmodes of the dimers with nanometric gaps.
We remark that the surface and volume responses act synergistically and shouldn't be understood solely separately since the presence of the QSRs also modifies the field profiles in the volume.

\begin{figure}[b!]
\centering
\includegraphics[width=0.43\textwidth]{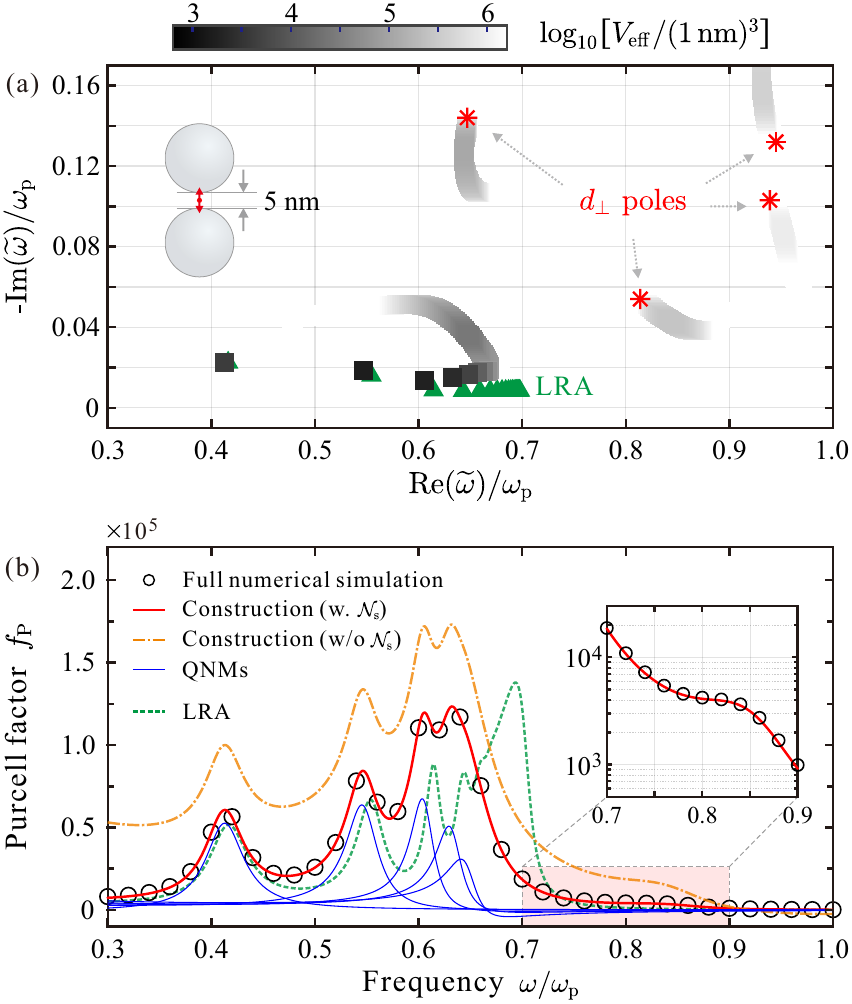}
\caption{The QNM analysis of emission enhancement for a dipolar emitter at the center of a Na nanosphere dimer ($60$ nm diameter and $5$ nm gap).
(a) The QNM spectrum calculated with the QSRs (gray squares) and under LRA (green triangles).
The gray scale is weighted by the effective mode volumes ($V_\mathrm{eff}$) evaluated at the dipole position.
The red asterisks denote the poles of $d_\perp(\widetilde\omega)$.
(b) The QNM construction of the Purcell factor spectra.
The thin blue traces denote the individual contributions of the QSR informed QNMs.
The construction of the total spectrum by $\sim$300 QNMs is plotted with the red trace and benchmarked with the QSR informed full numerical simulation denoted by black circles.
The green dashed and orange dash-dotted traces depict the spectra calculated under LRA and constructed with the QNMs normalized by only $\mathcal{N}_\mathrm{v}$, respectively.
}
\label{Fig:3}
\end{figure}

\section{Modal analysis of spontaneous emission enhancement}

Informed about the properties of the eigenmodes we are enabled to better interpret the optical responses of a nanosystem. One of the most enticing promises by nanoplasmonics is the huge spontaneous emission enhancement for a quantum emitter. We thus apply our QSR informed QNM theory to investigate the emission enhancement of a dipolar emitter placed at the gap center of the previous Na dimer with a 5 nm separation. To accurately delineate the enhancement spectrum, the full QNM landscape of the dimer is required. We carry out a thorough mode search and present in Fig.\,\ref{Fig:3}(a) the exhaustive QNM spectrum.
The QNMs by classical LRA calculation are also indicated with green triangles. The QSR informed QNM spectrum becomes considerably more intricate than the classical counterpart, where only one branch of QNMs exists. The corresponding branch now bends back to lower frequencies and opens up the possibility to mix with low-order QNMs. Besides, several new branches appear due to the multi-pole disperion of $d_\perp$ as indicated by the red asterisks. According to \eqref{Eq:fPm}, the contribution of a QNM to emission enhancement correlates with its mode volume. Therefore we paint the squre markers in grayscale to highlight their relevance. A darker color means a smaller mode volume and hence a larger contribution.
Apart from the lowest-order QNMs in almost black colors, we note those in the new branches shouldn't be omitted.

Given the numerically calculated QNMs we can determine analytically their individual contributions and construct the total spontaneous emission enhancement spectrum, \textit{i.e.}, the Purcell factor spectrum \footnotemark[1]. 
Since the emitter is only few nanometers from the metal surface, a large number of QNMs can be excited. The total spectrum converges to the red trace in Fig.\,\ref{Fig:3}(b) when about 300 QNMs are counted in. As a benchmark, the Purcell spectrum by full numerical calculation under $d$-parameter description is plotted with black circles. Excellent agreement between the analytical QNM construction and numerical calculation is witnessed. Albeit so many QNMs involved, the first 5 QNMs or $\mathrm{Q}_1$ to $\mathrm{Q}_5$ are sufficient to qualitatively account for the resonance features. The modal Purcell spectra of $\mathrm{Q}_1$ to $\mathrm{Q}_5$ are displayed in blue traces. $\mathrm{Q}_1$ to $\mathrm{Q}_4$ dominantly contribute to the four resonances of the total Purcell spectrum. Here the classical Purcell spectrum is also depicted in the green dashed trace for comparison. The redshift of the QSR informed QNMs and the increasing shifts with mode order are vividly observed. The classical counterpart of $\mathrm{Q}_5$ can also be identified in the classical Purcell spectrum, but the shoulder around $0.66\omega_\mathrm{p}$ disappears in the QSR informed Purcell spectrum.
That's because $\mathrm{Q}_5$ suffers extra damping due to the QSRs and its modal Purcell spectrum is severely broadened.
Aside from the four prominent resonances, the Purcell spectrum has a shoulder structure around 0.8$\omega_\mathrm{p}$ as shown in the zoomed-in view in the inset of Fig.\,\ref{Fig:3}(b).
It has no classical counterpart and purely originates in a new QNM branch. Moreover we underlie the crucial effect of the surface response by evaluating the Purcell spectrum with $\mathcal{N}_\mathrm{s}$ removed from the normalization factor. The resulting spectrum plotted in the orange dash-dotted trace is seen to considerably overestimate the Purcell factor.

\begin{figure}[tb!]
\centering
\includegraphics[width=0.45\textwidth]{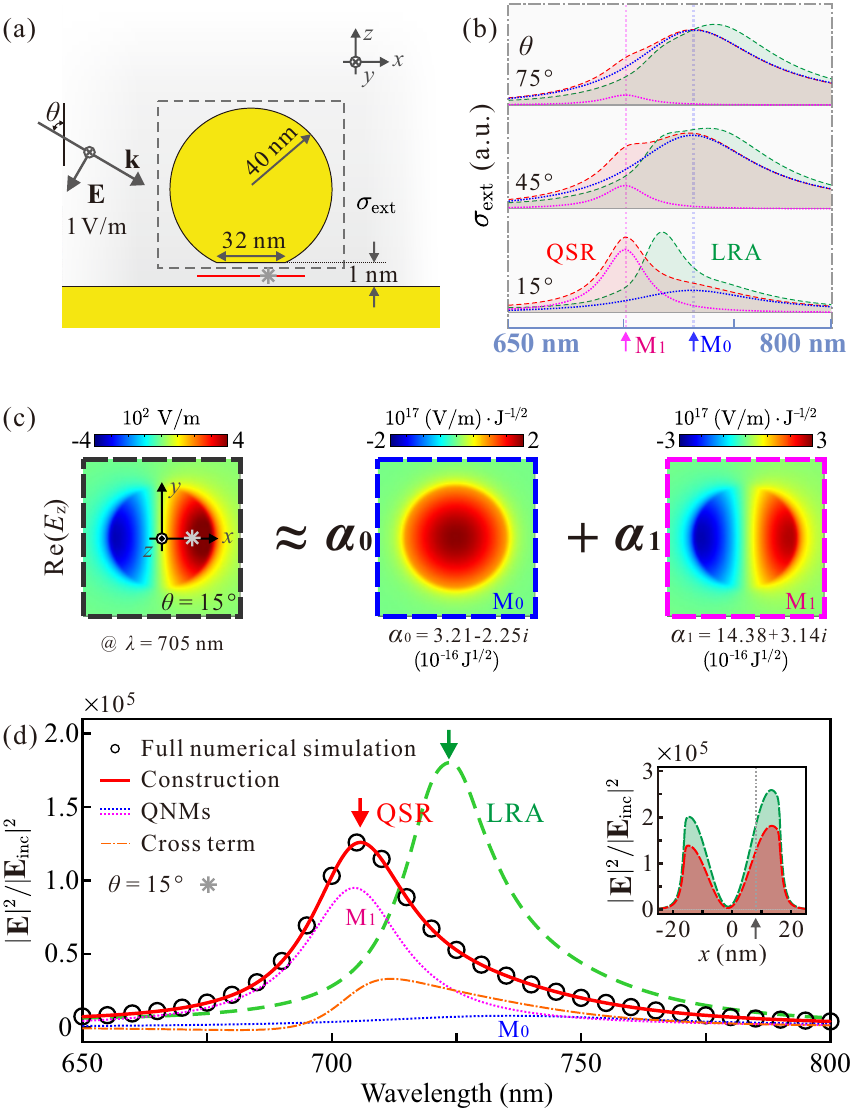}
\caption{The QNM analysis for a gold nanoparticle on mirror (NPoM) under the illumination of a plane wave.
(a) The schematic of the NPoM structure.
(b) The extinction cross section spectra for various incidence angles are calculated with the QSRs (red) and under LRA (green).
The blue and magenta dashed lines denote the respective contributions from the dominant QNMs $\mathrm{M}_{0,1}$.
(c) For the $\theta=15^\mathrm{o}$ incidence at $\lambda=705$ nm, $\mathrm{Re}\{E_z\}$ is constructed with $\mathrm{Re}\{\widetilde{E}_z\}$ of $\mathrm{M}_{0,1}$ on the cut plane at gap center;
\textit{cf.}\ the red line segment in (a).
(d) The QNM construction of light intensity enhancement at $(x,y)=(8,0)$ nm on the cut plane;
\textit{cf.}\ the gray asterisks in (a) and (c).
The spectra constructed by 8 QNMs, obtained by the QSR informed and classical LRA full numerical simulations are plotted with the red trace, black circles and green dashed trace, respectively.
The thin blue, magenta and orange traces denote the respective contributions from $\mathrm{M}_{0,1}$ and the cross coupling.
The inset shows the intensity enhancement along the line $y=0$ nm with the QSRs (red, $\lambda=705$ nm) or under LRA (green, $\lambda=723$ nm).}
\label{Fig:4}
\end{figure}

\section{Modal analysis of the responses to far-field excitation}
A majority of nano-optical processes in the end need to be probed with far-field optical excitation. In this section we demonstrate the application of the QSR informed QNM theory to a nanosystem under far-field excitation. Nanoparticle on mirror (NPoM) is an easy-to-fabricate and versatile nano-optical platform \cite{Baumberg2019ExtremeNanophotonics}. When the gap shrinks to a few nanometers, non-classical effects are bound to arise.
Concretely a gold NPoM structure with the geometry specifications given in Fig.\,\ref{Fig:4}(a) is illuminated by an oblique plane wave.
The cylindrical symmetry of the structure allows us to use a 2.5D numerical technique to conveniently calculate the QNMs \footnotemark[1].
The QNMs can be classified with azimuthal order $m$. For the plane wave incidence mainly the $m=0$ or $\mathrm{M}_0$ and two degenerate $m=\pm1$ modes are relevant. In practice, only $\mathrm{M}_0$ and a superposition of the $m=\pm1$ modes or $\mathrm{M}_1$ are excited by the plane wave \footnotemark[1].
Dependent on the incidence angle $\theta$ the two QNMs are excited with varying weights. Utilizing our QSR informed QNM theory, the two modes' contributions to extinction spectrum can be computed analytically and their sum gives a good estimation of the total extinction spectrum \footnotemark[1].
The spectra for $\theta=75^\mathrm{o}$, $45^\mathrm{o}$ and $15^\mathrm{o}$ are illustrated in Fig.\,\ref{Fig:4}(b).
$\mathrm{M}_0$/$\mathrm{M}_1$ is seen dominantly excited as the plane wave approaches the grazing/normal incidence.
The extinction spectra under classical LRA are also shown with green shaded areas.
As expected for Au, the QSR informed QNMs are blueshifted and $\mathrm{M}_1$ experiences a larger shift.

In the NPoM structure, the electric field confined in the nanogap is greatly enhanced. The electric field similarly has contributions from $\mathrm{M}_{0,1}$. Our QNM theory also allows a modal analysis of the near field profile. We exemplify the field construction for the $\theta=15^\mathrm{o}$ incidence at $\lambda=705$ nm. The expansion coefficients $\alpha_{0,1}$ of $\mathrm{M}_{0,1}$ at the excitation conditions can be determined using \eqref{QNMexpansion}. Then the electric field is simply constructed as $\mathbf{E}=\alpha_0\widetilde{\mathbf{E}}_0+\alpha_1\widetilde{\mathbf{E}}_1$.
The construction is visually illustrated in Fig.\,\ref{Fig:4}(c) with $\mathrm{Re}\{E_z\}$ on the cut plane at the gap center.
The asymmetric distribution due to the oblique incidence apparently results from the superposition of the symmetric profiles of $\mathrm{M}_{0,1}$.
Considering the 1 V/m amplitude of the incidence, several hundreds folds of enhancement in $E_z$ can be inferred from the leftmost map in Fig.\,\ref{Fig:4}(c). Enhancement in light intensity, including all components of the electric field, is more informative.
Yet light intensity is proportional to $|\mathbf{E}|^2$ and can't be decomposed into just isolated modal contributions as the electric field.
Cross coupling terms must be included. For the Au NPoM structure the cross coupling term between $\mathrm{M}_{0,1}$ should be considered.
In Fig.\,\ref{Fig:4}(d) we showcase the construction of light intensity enhancement evaluated at the position indicated by the asterisks in Fig.\,\ref{Fig:4}(a) and \ref{Fig:4}(c). Besides the two spectra due to $\mathrm{M}_{0,1}$, there is the cross coupling contribution depicted with the red dash-dotted line.
Here for $\theta=15^\mathrm{o}$, $\mathrm{M}_1$ dominates the enhancement around $\lambda=705$ nm.
Although $\mathrm{M}_0$ itself contributes little, the cross coupling contribution is significant and peaks between the resonances of $\mathrm{M}_{0,1}$.
A more accurate construction of the total enhancement spectrum is achieved when the contributions from 8 QNMs and their mutual couplings are added up.
It conincides with the QSR informed full numerical simulation (black circles). Compared with the prediction based on classical LRA (green dashed), non-classical effects clearly rectifies the resonance position and overestimated peak value. Nevertheless the LRA prediction regarding the spatial dependence of light intensity is seen acceptable as shown in the inset of Fig.\,\ref{Fig:4}(d). At their respective resonant wavelengths, the intensity distributions have two lobes and a nadir slightly displaced from the center.

\section{Conclusions}

In this work we have developed a quasinormal mode theory for nanometer scale electromagnetism with quantum surface responses. By introducing auxiliary variables for the surface responses, we have formulated the source-free Maxwell's equations into a composite linear eigenvalue problem consisting of inter-coupled volume and surface components. We've established an orthonormal relation resorting to the generalized unconjugated Lorentz reciprocity theorem. The orthonormal relation possesses a surface contribution, which results from a nonvanishing integral on the surface owing to the discontinuous boundary conditions. This surface contribution in the mode normalization factor provides an indicator for quantifying the significance of the quantum surface responses. Enabled by the orthogonal relation, we have formalized arbitrary external excitations and proposed analytical expansions of system responses by the quasinormal modes. The quasinormal mode theory has been thoroughly validated with analytical results and full numerical simulations (see \suppref{}), and proven highly accurate. We remark by passing our theory can be extended to encompass the quantum surface responses due to $d_\parallel$ (see \suppref{}). The present theory has substantially augmented the latest general framework for nanoscale electromagnetism by entailing \textit{e.g.}, insightful modal analysis and analytical prediction of non-classical nano-optical responses. Thus we envision the theory as a useful tool would serve well a wide range of areas in nano-optics, such as integrated nanophotonics \cite{Sun2015}, light-matter interaction in nanocavities \cite{Lalanne2018LigntInteraction}, Raman spectroscopy \cite{ZhiYuan2020RamanSpectroscopy} and surface physics \cite{Echarri2020}.

\begin{acknowledgments}
We would like to thank Wei Yan for helpful discussions. 
The numerical simulations in this paper are supported by the public computing service platform provided by the Network and Computing Center of HUST.
We acknowledge financial support from the National Natural Science Foundation of China (Grant Number 11874166).
\end{acknowledgments}

\footnotetext[1]{See Supplemental Material for details.}

\vspace*{.1cm}
\bibliographystyle{apsrev4-2}
\bibliography{text-main-v9.bib}
\end{document}